\documentclass[]{article}
\usepackage{amsmath}
\usepackage{wasysym}
\usepackage{latexsym}
\usepackage{amssymb}
\usepackage{bbm}
\usepackage{amsthm}
\usepackage{graphicx}
\usepackage{cite}   

\newcommand{\soq}{{SO}(4)}

\newcommand{\rd}[1]{\mathcal{#1}}

\newcommand{\eqa}{\begin{eqnarray}}
\newcommand{\neqa}{\end{eqnarray}}
\newcommand{\be}{\begin{equation}}
\newcommand{\ee}{\end{equation}}
\newcommand{\no}{\nonumber\\}
\def\f{\frac}

\newcommand{\ket}[1]{|{#1}\ra}
\def\ra{\rangle}
\def\la{\langle}

\newcommand{\bra}[1]{\la {#1}|}

\theoremstyle{plain}

\theoremstyle{definition}

\theoremstyle{remark}

\theoremstyle{plain}

\newcommand{\Tr}{\mathrm{Tr}}

\newcommand{\Hil}{\mathcal{H}}

\begin{document}
\title{The volume operator in covariant quantum gravity}
\author{You Ding, Carlo Rovelli \\[1mm]
\normalsize \em CPT%
\footnote{Unit\'e mixte de recherche (UMR 6207) du CNRS et des Universit\'es
de Provence (Aix-Marseille I), de la Meditarran\'ee (Aix-Marseille II) et du Sud (Toulon-Var); laboratoire affili\'e \`a la FRUMAM (FR 2291).} , CNRS Case 907, Universit\'e de la M\'editerran\'ee, F-13288 Marseille, EU}
\date{\small\today}
\maketitle\vspace{-7mm}
\begin{abstract}

\noindent A covariant spin-foam formulation of quantum gravity has been recently developed, characterized by a kinematics which appears to match well the one of canonical loop quantum gravity. In particular, the geometrical observable giving the \emph{area} of a surface has been shown to be the same as the one in loop quantum gravity.  Here we discuss the \emph{volume} observable. We derive the volume operator in the covariant theory, and show that it matches the one of loop quantum gravity, as does the area.   We also reconsider the implementation of the constraints that defines the model: we derive in a simple way the boundary Hilbert space of the theory from a suitable form of the classical constraints, and show directly that all constraints vanish  \emph{weakly} on this space.

\end{abstract}

\section{Introduction}
The spinfoam formalism \cite{Reisenberger:1994aw,spinfoam,spinfoam11,BC1,BC2,spinfoams} and
canonical loop quantum gravity (LQG) \cite{alrev,libro,lqg} can ideally be viewed as the covariant and the canonical versions, respectively,  of a background-independent quantum theory of gravity \cite{Rovelli:2010wq}. This scenario is nicely realized in three dimensions \cite{alex}, and there are recent attempts to implement it in quantum cosmology \cite{Ashtekar:2009dn,sfc}.  An important step ahead towards the realization of this scenario in the complete four dimensional theory has been taken with the recent introduction of two spin-foam models whose kinematics appears to match the one of LQG rather well, which we refer to as the new model  \cite{epr1,EPR,EPRL,P,LS2} and the Freidel-Krasnov-Livine-Speziale (FKLS) model \cite{LS1,FK}. The \emph{kinematics} of canonical loop quantum gravity, indeed, is rather well understood; in particular, the properties of the geometrical operators, including the area and the volume operators\cite{RS,area,volume} are well established. (On the volume operator, see also \cite{semi}.)    The area operator of the new spinfoam model has been derived in \cite{EPRL,EP} and shown to match the LQG one.  Does the volume do so as well?

The volume observable in the covariant spinfoam language has not been constructed yet. The essential property of the volume operator is that it has contribution only from the nodes of a spin network state. Thus the only possible action of the volume operator is on the intertwiners. That's the reason why there is no generic well-defined volume operator in the old Barrett-Crane (BC) model\cite{BC2}, based on the vertex amplitude introduced by Barrett and Crane \cite{BC1}, where intertwiners are fixed. In fact, the absence of the volume operator in the Barrett-Crane theory can be traced precisely to the key problem of the BC model: the fact that \emph{intertwiner} quantum numbers are fully constrained.  This follows from the $SO(4) \rightarrow SU(2)$ gauge fixing and the way certain second class constraints are imposed, arguably incorrectly, strongly.  The new model \cite{epr1,EPR,EPRL,P} imposes second class simplicity constraints weakly, rather than strongly as in the BC theory.  This choice frees intertwiner degrees of freedom, and the volume operator can be nontrivial. In this model, the state space where the constraints vanish weakly turns also out to match that of LQG, providing a solution to the problem of connecting the covariant $SO(4)$ spinfoam formalism with the canonical $SU(2)$ spin-network one. To complete this identification, we construct here the volume operator in the covariant spinfoam picture, and show explicitly that it matches the corresponding LQG canonical operator.

As a first step, in the next section we review the derivation of the boundary state space of the spinfoam theory. We do so for completeness, but also in order to clean up and simplify previous derivations in the literature.  In particular, we show explicitly and directly that all the constraints vanish in a weak sense in the physical boundary space.  The form of the constraints that we write turns out to strongly simplify the study of the volume operator.

We work only in the Euclidean theory, on a fixed triangulation, and assume here that the Barbero-Immirzi parameter $\gamma$ is positive. The paper is organized as follows. In section 2, we review the definition of the physical boundary Hilbert space. The volume operator is constructed and shown to match the LQG operator in Section 3.

\section{The boundary space}
\subsection{Classical theory}
Following \cite{EPR,EPRL}, we start with a Regge geometry \cite{regge}
on a fixed triangulation. Consider a 4d triangulation, which is formed by oriented
4-simplices, tetrahedra, triangles, segments and points. We call
$v,t$ and $f$ respectively the 4-simplices, the tetrahedra and the
triangles of the triangulation. For each simplex $v$, we introduce a variable $e_{\mu}^I(v)$: a right-handed tetrad one-form, constant over a coordinate patch covering the simplex $v$, with the determinant $\mathrm{det}(e)>0$ positive. Here $\mu=(0,a)$
and $a=1,2,3$ are spacetime indices, while $I=(0,i)$
and $i=1,2,3$ are internal indices (the value 0 instead of 4 is for later convenience and does not indicate a Lorentzian metric).
Without loss of generality, we can choose a linear coordinate system with basis vectors $\overrightarrow{X}_{\mu}$ parallel with four edges of $v$ emanating from the same point, and where the (coordinate) length of the four segments is 1.  Consider in particular the tetrahedron $t$ spanned by the three
vectors  $\overrightarrow{X}_a$. To each triangle $f_a$
(coordinate-)normal to the coordinate basis vector
$\overrightarrow{X}_a$, we associate a bivector
${}^*B_a(t)$ defined by: %
\be %
{}^*B^{IJ}_a=\f{1}{2}\epsilon_a^{\ bc}e^I_be^J_c.%
\label{B}
\ee
$B_f(t)$ can be seen as elements in the algebra $\mathfrak{g}=\mathfrak{so}(4)$, in the Euclidean case, and  $*$ stands for the Hodge dual in the internal indices.
If we  choose $B_f(t)$ as independent variables instead of the
tetrads, and $n_I$ denotes the normal to the tetrahedron $t$, the
simplicity constraints on $B_f(t)$, which assure that a tetrad field exist, can be stated as follows \cite{EPR,EPRL}:
\be
C_{f}^J:=n_I\ ({}^* B_f(t))^{IJ} = 0. \label{c2}
\ee
The usual quadratic diagonal
\be
C_{ff}:={}^* B_f(t)\cdot B_f(t) = 0
\label{c1}
\ee
and off-diagonal
\be
C_{ff'}:={}^* B_f(t)\cdot B_{f'}(t) = 0
\label{c0}
\ee
simplicity constraints can be easily shown to follow from (\ref{c2}).
Here the dot stands for the scalar product in the $\mathfrak{so}(4)$ algebra.
In addition, we should impose the closure constraint
\begin{equation}\sum_{f \in  \partial t} B_f(t) = 0.
\label{c3}
\end{equation}
The new linear simplicity constraint (\ref{c2}) selects the solution of the quadratic constraints where $B_f=\int_f{}^*(e\wedge e)$. This reformulation is central for the new model \cite{epr1,EPR,EPRL,P}.
In particular, if we choose a ``time" gauge where $n^I=(0,0,0,1)$, the
 simplicity constraint (\ref{c2}) turns out to be
\begin{eqnarray}
{}^*B^{0i}_f(t) = 0\label{c2g}.
\end{eqnarray}

The classical discrete action is \cite{EP,EPRL}
\begin{align}
\nonumber
S=&-\sum_{f \in int\Delta} \Tr\left[B_f(t)U_f(t)
+ \frac{1}{\gamma}\,^* B_f(t) U_f(t) \right]  \\
\label{discaction}
&-\sum_{f \in \partial\Delta} \Tr\left[B_f(t)U_f(t,t')
+ \frac{1}{\gamma}\,^* B_f(t) U_f(t,t') \right],
\end{align}
where $U_f(t, t')$ is the group element of $SO(4)$,
giving the parallel transport across each triangle $f$ bounding $t$
and $t'$ and $U_f(t):= U_f(t,t)$ is the holonomy around the full link,
starting at $t$. We use here unites where $2\kappa = 16\pi G=1$ and $\gamma$  the Barbero-Immirzi parameter.  This action, plus the simplicity and closure constraints defines a discretization of general relativity \cite{EPR,EPRL}. From the action, we can read off the boundary variables as
$B_f(t) \in \mathfrak{so}(4)$, $U_f(t,t') \in SO(4)$.
One can also see that the variable conjugate to $U_f(t,t')$ is
\begin{align}
J_f(t):=&B_f(t)+\f{1}{\gamma}{}^*B_f(t),
\end{align}
inverting which gives
\begin{align}
\,^*B_f(t)=\frac{\gamma^2}{1-\gamma^2}\Big(\frac{1}{\gamma}J_f(t)-^*J_f(t)\Big)\label{inver}.
\end{align}

Thus to each boundary triangle $f$  in the boundary of the triangulation, we have an $SO(4)$ group element $U_f$ and,
as conjugate variable an $\mathfrak{so}(4)$ algebra element $J_f$. It is convenient to think these variables as associated with the
links of the graph formed by the one-skeleton of the cellular complex dual to the boundary triangulation. Notice
that these define precisely the same boundary phase space as the one of an $SO(4)$ lattice Yang-Mills theory. As in
Yang-Mills theory, the symplectic structure can be taken to be \cite{EPR}
\begin{align}
\{ U_f ,U_{f'} \} &=0 , \nonumber \\
\{ (J_f)^{IJ} ,U_{f'} \} &=\delta_{ff'}\  U_f\ \tau^{IJ} ,\\
\{ (J_f)^{IJ} ,  (J_f')^{KL}  \} &= \delta_{ff'}\
\lambda^{IJ\,KL}_{MN} (J_f)^{MN}\nonumber,
\end{align}
where $\tau^{IJ}$ and $ \lambda^{IJ\,KL}_{MN} $ are, respectively,
the generators and the structure constants of $SO(4)$.

In terms of the momentum variable $J_f$, the constraints (\ref{c2}) and (\ref{c3}) read respectively:
\begin{align}
 &C_f^J=n_I\left((^*
J_f)^{IJ}-\frac{1}{\gamma}J_f^{IJ}\right) = 0 \label{CJ},\\
&\sum_{f \in \partial t} J_f(t) =  0.\label{sumJ}
\end{align}

For the gauge-fixed version, introduce $L_f^j:=\frac{1}{2}\epsilon^{j}{}_{kl}J_f^{kl}$ and $K_f^j:=J_f^{0j}$ , which are respectively the generators of the $SO(3)$ subgroup that leaves $n_I$ invariant, and the generators of the corresponding boosts. Then the simplicity constraint (\ref{c2}) becomes simply
\be
          K^i_f=\gamma\ L^i_f.
          \label{key}
\ee
This is the key constraint.
In terms of $(K_f^i,\,L_f^i)$, the closure constraint (\ref{sumJ}) turns out to be
\begin{subequations}\label{closure}
\begin{align}
\sum_{f\in\partial t}L_f^i&=0,\label{closureL}\\
\mathrm{and}\quad\sum_{f\in\partial t}K_f^i&=0.\label{closureK}
\end{align}
\end{subequations}
If we further make the self-dual/anti-self-dual decomposition of $J_f^{IJ}$:
\begin{align}
 J_f^{(\pm)i}:=&\f{1}{2}(L_f^i\pm
K_f^i),\label{JLK}
\end{align}
the simplicity constraint (\ref{CJ}) implies
\begin{align}
&C^i_{f}=(1-\gamma)J_f^{(+)i}-(1+\gamma)J_f^{(-)i}\label{co2}.
\end{align}
In terms of $J^{(\pm)i}$, the usual quadratic diagonal simplicity constraint (\ref{c1}), which follows from the new simplicity constraint (\ref{c2}) or (\ref{co2}), can be reexpressed as
\begin{align}
&C_{ff}=(1-\gamma)^2J_f^{(+)2}-(1+\gamma)^2J_f^{(-)2}\label{co1}.
\end{align}

\subsection{Quantization}
From the discrete boundary variables and their symplectic
structure, we can construct the Hilbert space associated
with a boundary or 3-slice.
To do this, it is simpler to
switch to the dual, 2-complex picture, $\Delta^*$.
For each 3-surface $\Sigma$ intersecting no vertices of $\Delta^*$,
let $\gamma_{\Sigma} := \Sigma \cap \Delta^*$.  The Hilbert space
associated with $\Sigma$ is then
\begin{equation}
\Hil_{\Sigma} = L^2\left({Spin}(4)^{|L(\gamma_\Sigma)|},\mathrm{d}\mu_{\mathrm{Haar}}\right),
\end{equation}
where we replace $SO(4)$ with its covering group $Spin(4)=SU(2)\times SU(2)$ and  $\mu_{\mathrm{Haar}}$ is the Haar measure on the group $Spin(4)$; $|L(\gamma_\Sigma)|$ denotes the number of links in
$\gamma_\Sigma$. Let $\hat{J}_f(t)^{IJ}$ denote the
right-invariant vector fields, determined by the basis $J^{IJ}$ of
$\mathfrak{su}(2)\oplus\mathfrak{su}(2)$, on the copy of $Spin(4)$ associated with the link
$l = f \cap \Sigma$ determined by $f$, with orientation such that
the node $n = t \cap \Sigma$ is the source of $l$.

By Peter-Weyl theorem, $\Hil_{\Sigma}$ can be decomposed as follows
\begin{equation}
\Hil_{\Sigma} = \bigoplus_{j_l}\bigotimes_l \left(\Hil_{j_l}^* \otimes \Hil_{j_l}\right),
\end{equation}
where $j_l$ is an assignment of a $Spin(4)$ representation to each link $l$ and $\Hil_{j}$ is the carrier space of the representation $j$.  The two Hilbert spaces associated to the link $l$ are naturally associated to the two nodes that bound the link $l$, because they transform under the action of a gauge transformation at one end of the link. Regrouping the four Hilbert spaces associated to each node $n$, the last equation can be rewritten in the form
\begin{equation}
\Hil_{\Sigma} = \bigoplus_{j_l}\bigotimes_n  \Hil_{n}.
\end{equation}
Here the Hilbert space associated to a node $n$ is
\begin{equation}
\Hil_{n} = \bigotimes_{a=1}^4  \Hil_{j_{a}},
\end{equation}
where $a=1,2,3,4$ runs here over the four edges that join at the node $n$ (that is, the four faces of the boundary tetrahedron),
and we have identified the Hilbert space carrying a representation and its dual.
 We restrict our attention to a single boundary tetrahedron $t$, and its associated Hilbert space $\Hil_{n}$, which we call simply $\Hil$ in the following.

The irreducible unitary representations of $Spin(4)$ are labelled by a couple of spins $(j^+,j^-)$ and are given by the tensor product of two $SU(2)$ irreducibles. That is $\Hil:=\Hil_n$ has the structure
\be
\Hil = \bigotimes_{a=1}^{4}\Hil_{(j_a^+,j_a^-)}=\bigotimes_{a=1}^{4}\big(\Hil_{j_a^+}\otimes\Hil_{j_a^-}\big).
\ee
The physical intertwiner state space $\mathcal{K}_{\mathrm{ph}}$ is a subspace of this space, where the constraints hold in a suitable sense.

As a first step to impose the constraints, let us restrict the representations to the ones that satisfy
\be%
j^+=\Big|\frac{1+\gamma}{1-\gamma}\Big|j^-,
\label{gammasimple}
\ee%
which satisfies the usual quadratic diagonal simplicity constraint (\ref{co1}) in the classical limit, and is what we need to recover the correct classical theory in the limit. We call $\gamma$-simple the $Spin(4)$ representations that satisfy this relation.

Next, the Clebsch-Gordan decomposition for the single component of $\Hil$ associated with a single boundary face $f$ gives
\begin{equation}
\mathcal{H}_{j^+ \otimes j^-} = \mathcal{H}_{j^+} \bigotimes
\mathcal{H}_{j^-} = \bigoplus_{p=|j^+-j^-|}^{j^++j^-}\mathcal{H}_p.\label{CG}
\end{equation}
Consider the highest spin term in each factor for $\gamma<1$ and the lowest for $\gamma>1$ respectively; this selects the ``extremum'' subspace
\begin{align}
\mathcal{H}^{\mathrm{max}}&=\bigotimes_{a=1}^4 \mathcal{H}_{j^++j^-},\quad \mathrm{for}\ \gamma<1;\\
\mathcal{H}^{\mathrm{min}}&=\bigotimes_{a=1}^4 \mathcal{H}_{j^+-j^-},\quad \mathrm{for}\ \gamma>1.
\end{align}
We are now going to show that in this space (with (\ref{gammasimple}) holding), the simplicity constraint (\ref{co2}) is satisfied weakly. That is, the action of the constraints on the states in
${\Hil}^{\mathrm{ext}}$ results in states orthogonal to
${\Hil}^{\mathrm{ext}}$. Namely, $\bra{\Psi}\hat{C}^{i}_{f}\ket{\Phi}=0,\ \forall \,\Psi,\,\Phi\,\in{\Hil}^{\mathrm{ext}}$. This follows from the following considerations. 
$\forall \Psi,\,\Phi\,\in {\Hil}^{\mathrm{ext}}$, consider the matrix element of the form (\ref{co2}) of the simplicity constraints
\be
\bra{\Psi}\hat{C}^{i}_{f}\ket{\Phi}=(1-\gamma)\bra{\Psi}\vec{{J}_f^+}\ket{\Phi}-(1+\gamma)\bra{\Psi}\vec{{J}_f^-}\ket{\Phi}\label{mepm}
\ee
and write the r.h.s. of this equation in a representation where elements of $\Hil_j$ are symmetric spinors with $2j$ indices. The generators of $SU(2)$ are then Pauli matrices $\sigma^A_i{}_B$  acting on each index. {For} $\gamma<1$,
\begin{align}\label{proof}
\bra{\Psi}\hat{C}^{i}_{f}\ket{\Phi}
=&(1-\gamma)\Psi_{(A_1...A_{2j^+}B_1...B_{2j^-})}\sum_{p=1}^{2j^+}\sigma^{A_p}_i{}_{\widetilde{A}_p}\Phi^{(A_1...\widetilde{A}_p...A_{2j^+}B_1...B_{2j^-})}\no
&-(1+\gamma)\Psi_{(A_1...A_{2j^+}B_1...B_{2j^-})}\sum_{p=1}^{2j^-}\sigma^{B_p}_i{}_{\widetilde{B}_p}\Phi^{(A_1...A_{2j^+}B_1...\widetilde{B}_p...B_{2j^-})}\no
=&2\big((1-\gamma)j^+-(1+\gamma)j^-\big)\Psi_{(A_1...A_{2j^+}B_1...B_{2j^-})}\sigma^{A_1}_i{}_{\widetilde{A}_1}\Phi^{(\widetilde{A}_1...A_{2j^+}B_1...B_{2j^-})}\no
=&0.
\end{align}
The first step is obtained by the symmetry of the highest spin states, and the last follows from (\ref{gammasimple}).  Therefore the simplicity constraint is implemented weakly in ${\Hil}^{\mathrm{max}}$ for $\gamma<1$. For $\gamma>1$, the state $\ket{\Psi}$ in $\Hil^{\mathrm{min}}$ can be expressed as $\Phi^{(A_1...A_{2j^+})(B_1...B_{2j_-})}=\epsilon^{A_1B_1}...\epsilon^{A_{2j^-}B_{2j^-}}\Phi^{(A_{2j^-+1}...A_{2j^+})}$, on which the action of $J^{(-)}$ can be obtained as
\begin{align}
\vec{J}^{-}\Phi^{(A_1...A_{2j^+})(B_1...B_{2j_-})}=&\sum_{p=1}^{2j^-}\sigma_i^{B_p}{}_{\widetilde{B}_p}\epsilon^{A_1B_1}...\epsilon^{A_p\widetilde{B}_p}...\epsilon^{A_{2j^-}B_{2j^-}}\Phi^{(A_{2j^-+1}...A_{2j^+})}\nonumber\\
=&-\sum_{p=1}^{2j^-}\sigma_i^{A_p}{}_{\widetilde{A}_p}\epsilon^{A_1B_1}...\epsilon^{\widetilde{A}_pB_p}...\epsilon^{A_{2j^-}B_{2j^-}}\Phi^{(A_{2j^-+1}...A_{2j^+})}\nonumber\\
=&-\sum_{p=1}^{2j^-}\sigma_i^{A_p}{}_{\widetilde{A}_p}\Phi^{(A_1...\widetilde{A_{p}}...A_{2j^+})(B_1...B_{2j_-})}.
\end{align}
Hence the matrix elements of the simplicity constraint can be obtained as
\begin{align}
\bra{\Psi}\hat{C}^{i}_{f}\ket{\Phi}&=2\big((1-\gamma)j^++(1+\gamma)j^-\big)\Psi_{(A_1...A_{2j^+})(B_1...B_{2j^-})}\sigma^{A_1}_i{}_{\widetilde{A}_1}\Phi^{(\widetilde{A}_1...A_{2j^+})(B_1...B_{2j^-})}=0\nonumber
\end{align}
The last step follows again from (\ref{gammasimple}). Therefore the space $\Hil^{\mathrm{ert}}$ solves the simplicity constraint.
The physical intertwiner space associated with a single node $n$ is then obtained by solving the closure
constraint (\ref{closure}) weakly in the space ${\Hil}^{\mathrm{ext}}$, which turns out
to be
\be%
{\cal
K}_{\mathrm{ph}}= \mathrm{Inv}_{\mathrm{SU}(2)}[{\Hil}^{\mathrm{ext}}].
\label{inv}
\ee%
To show that the closure constraints (\ref{closure}) hold weakly on this space, observe the matrix elements (\ref{mepm}) of the simplicity constraint implies
\begin{align}
\bra{\Psi}K^{i}_{f}\ket{\Phi}=\gamma\bra{\Psi}L^{i}_{f}\ket{\Phi}.\label{meKL}
\end{align}
The l.h.s of (\ref{closureL}) is the generator of $SU(2)$ transformations at the node and vanishes strongly on (\ref{inv}) by definition; the l.h.s of (\ref{closureK}) is proportional weakly to the one of (\ref{closureL}) by (\ref{meKL}) and therefore vanishes weakly. Hence the $SU(2)$-invariant space turns out to be $SO(4)$-invariant space in the weak sense.
Thus $\mathcal{K}_{\mathrm{ph}}$ is the intertwiner space as a solution of \emph{all} the constraints: all the constraints hold weakly.

The total
physical boundary space  ${\rd H}_{\mathrm{ph}}$ of the theory is then
obtained as the span of spin-networks in $L^2[{Spin}(4)^L/{Spin}(4)^N,\mathrm{d}{\mu}_{\mathrm{Haar}}]$ with
$\gamma$-simple representations on edges and with intertwiners in the spaces
${\rd K}_{\mathrm{ph}}$ at each node.

We have then the remarkable result that $\mathcal{K}_{\mathrm{ph}}$ is naturally  isomorphic to the $SU(2)$ intertwiner space, and therefore  the
constrained boundary space ${\rd H}_{\mathrm{ph}}$ can be identified with the $SU(2)$
LQG state space  ${\rd H}_{{SU}(2)}$  associated to the graph  which is
dual to the boundary of the triangulation, namely the space of the
$SU(2)$ spin networks on this graph.

Since we have not proven that the physical Hilbert space considered is the \emph{maximal} space where the constraints hold weakly, one might worry that the physically correct quantization of the degrees of freedom of general relativity could need a larger space. Also, it has been pointed out that imposing second class constraints weakly might lead to inconsistencies in some cases\cite{Alexandrov:2010pg}. In the present case, however, these worries are not relevant, since the space obtained is directly related to the one of the canonical theory, which we can trust to capture the degrees of freedom of gravity correctly.

Let us now consider the geometrical operators in these two versions. Classically, the area $A(f)$ of a triangle $f$ is given by $A(f)^2={\frac{1}{2}({}^*B_f)^{IJ}\cdot({}^*B_f)_{IJ}}$. If we fix the time gauge,  we have
$A_3(f)^2={\frac{1}{2}({}^*B_f)^{ij}\cdot({}^*B_f)_{ij}}$. These two quantities are equal up to a constrained term.  As shown in \cite{EP,EPRL}, using the constraints, the operator related to $A_{3}(f)^2$ can be obtained as $A_3(f)^2=\kappa^2\gamma^2L_f^2$, which matches three-dimensional area as determined by LQG, including the correct Barbero-Immirzi parameter proportionality factor.  Let us now turn to study the volume operator on this space ${\rd H}_{\mathrm{ph}}$ and its relation with the $SU(2)$ volume in LQG.

\section{The Volume}

It is easy to see from the definition of $e_a^i(v)$ given at the beginning of the previous section, that the volume of the tetrahedron $t$ is given by
\be
V(t)=\f{1}{6}{\det (e(v))}.\label{volume}
\ee
In terms of the variables ${}^*B$ defined in (\ref{B}), the volume of a boundary tetrahedron $t$ reads
$V$
related to the tetrahedra $t$ as %
\be%
V(t)=\sqrt{\f{1}{27}\epsilon^{abc}\Tr[{}^*B_a{}^*B_b{}^*B_c]}\label{v4}
\ee%
To see this, let the gauge-fixed   simplicity constraint
(\ref{c2g}) hold, then the ${}^*B^{0i}_f(t)$ vanish and the above
quantity is equal to
\begin{align}
V_3(t)=\sqrt{\f{1}{27}\epsilon^{abc}{}^*B^{ij}_a{}^*B^{jk}_b{}^*B^{ki}_c}
=\f{1}{6}{\det (e)},\label{v3t}
\end{align}
which is exactly the expression (\ref{volume}) of the discrete volume.
Note that the $\soq$ volume $V_{\soq}(t)$ is gauge invariant, hence we can obtain eq (\ref{v4}) by the gauge-fixed version (\ref{v3t}) without loss of generality.
Going to the variables $J$, and using (\ref{inver}), the volume reads
\begin{align}
&V(t)=\sqrt{\f{1}{27}\Big(\frac{\gamma^2}{1-\gamma^2}\Big)^3\epsilon^{abc}\Tr\Big[\Big(\frac{1}{\gamma}J_a-^*J_a\Big)\Big(\frac{1}{\gamma}J_b-^*J_b\Big)\Big(\frac{1}{\gamma}J_c-^*J_c\Big)\Big]}
\label{v4j}
\end{align}

The volume operator $\hat{V}(t)$ of the tetrahedron $t$ is
 then formally given by (\ref{v4j}) with $J^{IJ}$ replaced by the corresponding operators:
\begin{align}
\hat{V}(t)=\sqrt{\f{1}{27}\Big(\frac{\gamma^2}{1-\gamma^2}\Big)^3\epsilon^{abc}\Tr\Big[\Big(\frac{1}{\gamma}\hat{J}_a-^*\hat{J}_a\Big)\Big(\frac{1}{\gamma}\hat{J}_b-^*\hat{J}_b\Big)\Big(\frac{1}{\gamma}\hat{J}_c-^*\hat{J}_c\Big)\Big]}.
\label{hatv}
\end{align}
However, the physical volume should be defined on the physical boundary space $\cal{H}_{\mathrm{ph}}$, satisfying the constraints.
Since the volume operator does not change the graph of the spin network sates, nor the coloring of the links, its action can be studied on the Hilbert space associated to a single node. Consider the matrix element of the square of the volume operator between two states in the physical Hilbert space (we drop the hats):
\be
\bra{i} V(t)^2 \ket{j}
=\f{1}{27}\big(\f{\gamma^2}{1-\gamma^2}\big)^3\epsilon^{abc}\bra{i} \big(\f{1}{\gamma}J_a^{ij}-{}^*J^{ij}_a\big)\big(\f{1}{\gamma}J_b^{jk}-{}^*J^{jk}_b\big)\big(\f{1}{\gamma}J_c^{ki}-{}^*J_c^{ki}\big) \ket{j}.
\ee
Writing this in terms of $L$ and $K$ components gives
\be
\bra{i} V(t)^2 \ket{j}
=\f{1}{27\cdot 8}\big(\f{\gamma^2}{1-\gamma^2}\big)^3\epsilon^{abc}\epsilon^{ij}_{\ \ m}\epsilon^{jk}_{\ \ n}\epsilon^{ki}_{\ \ p}\bra{i} \big(\f{1}{\gamma}L_a^{m}-K^{m}_a\big)\big(\f{1}{\gamma}L_b^{n}-K^{n}_b\big)\big(\f{1}{\gamma}L_c^{p}-K_c^{p}\big) \ket{j}.\label{simplify}
\ee
Notice that the intertwiner space is the subspace of the product of the space $\Hil_{a}$ associated to the link $a$, and the action of $(K_a,\,L_a)$ is in fact on $\Hil_a$. Hence we can use the form  (\ref{meKL}) of the simplicity constraint to simplify Eq. (\ref{simplify}), although the r.h.s seems a polynomial.
Using the form (\ref{meKL}) of the constraint, we can rewrite it as
\begin{align}
\bra{i} V(t)^2 \ket{j}
=&\f{1}{27\cdot 8}\big(\f{\gamma^2}{1-\gamma^2}\big)^3\big(\f{1}{\gamma}-\gamma\big)^3\epsilon^{abc}\epsilon_{ijk}\bra{i} L_a^{i}L_b^{j}L_c^{k} \ket{j}
\end{align}
and a little algebra gives
\be
\bra{i} V(t)^2 \ket{j}
=\gamma^3 \ \bra{i} \epsilon^{abc}\epsilon_{ijk}L_a^{i}L_b^{j}L_c^{k} \ket{j}.
\ee
That is
\be
V(t)
=\gamma^{\frac32} \sqrt{\epsilon^{abc}\epsilon_{ijk}L_a^{i}L_b^{j}L_c^{k}}
\ee
Now, the operator on the r.h.s. is precisely the LQG volume operator  $V_{\mathrm{LQC}}$, as it acts on ${\cal
K}_{\mathrm{ph}}$ including the correct dependence on the Barbero-Immirzi parameter $\gamma$.
\section*{Acknowledgments}
Thanks to Eugenio Bianchi for enlightening discussions. We wish also to thank Jonathan Engle, Simone Speziale, Franck Hellmann, Roberto Pereira and Muxin Han for helpful comments and suggestions. Y. D. is supported by CSC scholarship No. 2008604080.

\end{document}